\documentclass[lettersize,journal]{IEEEtran}
\usepackage{amsmath,amsfonts}
\usepackage{algorithmic}
\usepackage{algorithm}
\usepackage{array}
\usepackage[caption=false,font=normalsize,labelfont=sf,textfont=sf]{subfig}
\usepackage{hyperref}
\usepackage{multirow}
\usepackage{textcomp}
\usepackage{stfloats}
\usepackage{url}
\usepackage{verbatim}
\usepackage{graphicx}
\usepackage{cite}
\usepackage{makecell}
\usepackage{siunitx}
\usepackage{upgreek}
\usepackage{amssymb}

\usepackage{xcolor}
\definecolor{darkgreen}{rgb}{0.1,0.6,0.1}
\definecolor{steelblue}{rgb}{.273,.508,.703}

\newcommand{\Potassium}{$^{40}$K}
\newcommand{\Thorium}{$^{232}$Th}
\newcommand{\Uranium}{$^{238}$U}
\newcommand\mc[1]{\multicolumn{1}{c}{#1}}

\begin{document}

\title{Computed models of natural radiation backgrounds in qubits and superconducting detectors}

\author{Joseph Fowler, Ian Fogarty Florang, Nathan Nakamura, Daniel Swetz, Paul Szypryt, Joel Ullom~\IEEEmembership{National Institute of Standards \& Technology, Boulder, Colorado 80305 USA\vspace{-5mm}}
}

\markboth{IEEE Transactions on Applied Superconductivity}%
{Shell \MakeLowercase{\textit{et al.}}: A Sample Article Using IEEEtran.cls for IEEE Journals}


\maketitle

\begin{abstract}
Naturally occurring radiation backgrounds cause correlated decoherence events in superconducting qubits.
These backgrounds include both gamma rays produced by terrestrial radioisotopes and cosmic rays.  We use the particle-transport code Geant4 and the PARMA summary of the cosmic-ray spectrum to model both sources of natural radiation and to study their effects in the typical substrates used in superconducting electronics. We focus especially on three rates that summarize radiation's effect on substrates. We give analytic expressions for these rates, and how they depend upon parameters including laboratory elevation, substrate material, ceiling thickness, and wafer area and thickness. The modeled rates and the distribution of event energies are consistent with our earlier measurement of radiation backgrounds using a silicon thermal kinetic-inductance detector.
\end{abstract}


\section{Superconducting qubit or sensor backgrounds}

Quantum computers are now the subject of intensive research and massive commercial investments. They employ qubits based on a wide variety of systems; in many, the qubits are made of superconducting circuits deposited on silicon substrates~\cite{Bravyi2022,Kjaergaard2020}. Naturally occurring sources of background radiation can cause correlated decoherence of such qubits~\cite{Zmuidzinas2012,Cardani2021,Wilen2021, McEwen2022,Thorbeck2023}. Similarly, cryogenic quantum sensors like transition-edge sensors (TES) are subject to unwanted backgrounds arising from the same sources. While such backgrounds may be irrelevant to TES instruments used for signal-dominated measurements, they can be the primary limitation for rare-event searches, such as the attempted direct detection of dark matter.

The dominant sources of background radiation are cosmic rays produced in earth's atmosphere and gamma rays from radiogenic processes in building materials and bedrock. These sources are well understood. Their effects can be modeled with a variety of standard computational tools, but such modeling is a time-consuming process that generally requires specialized expertise. For the ultimate analysis of a critical experiment, this effort is easily justified; for other applications, however, a few approximate figures and basic scaling laws would suffice.

Unfortunately, there are few simple but quantitative estimates available to predict the impact of natural radiation backgrounds upon qubits or sensors made from semiconductor substrates in sizes typical of such circuits. We have developed such estimates as part of our investigations with a superconducting Thermal Kinetic-Inductance Detector (TKID)~\cite{fowler2024}. In this work, we attempt to capture the rates and the main scaling rules for the broadest possible applicability to users of superconducting qubits or radiation detectors. We base the results on a nominal cryogenic circuit: a silicon substrate \qty{500}{\micro\m} thick and \qty{10}{mm} square in a laboratory at sea level.

\section{Terrestrial and cosmic sources of background}

Gamma rays produced by naturally occurring radioactive materials are ubiquitous. All concrete and other building materials, and all bedrock and dirt contain traces of radioactive isotopes. Concrete foundations and the underlying earth present unavoidable sources of background radiation. The radioactive isotopes primarily responsible for gamma-ray emission are the primordial isotopes \Potassium, \Uranium, and \Thorium, and their various progeny among the actinide elements. \Potassium\ nuclei decay to stable elements, but \Uranium\ and \Thorium\ stand at the beginning of two long decay chains, each producing over a dozen radioactive isotopes and hundreds of individual nuclear gamma-ray and atomic x-ray emissions. Fortunately, the simplifying approximation of secular equilibrium between unstable isotopes in a chain can be used in most situations. That is, every unstable isotope but those at the top of a chain reach a balance with equal production and decay rates. Under this assumption, the specific activity (decay rate per kg of material) of the first unstable isotope describes the entire chain.

Radon, being gaseous, can move separately from the other elements in a geologic matrix. It is therefore possible for the portions of a decay chain that precede and follow radon to be out of equilibrium. Thus we allow for the specific activity of the two halves to be distinct. We name these half-chains \Uranium-a and \Uranium-b (and similarly for \Thorium).

The radioactivity level of concrete varies by factors of 5-10 among samples, depending on the source of the rock aggregate and of the chemical cement used to bind it~\cite{Papastefanou2005,Suzuki2000,Trevisi2012}. We base the nominal specific activity levels on the typical activity concentration found in common building materials in Europe~\cite{Kovler2012}. These values are \qty{400}{Bq.kg^{-1}}, \qty{30}{Bq.kg^{-1}}, and \qty{40}{Bq.kg^{-1}} for \Potassium, \Thorium, and \Uranium, respectively, as well as for each isotope in equilibrium with these parent isotopes.

Cosmic rays that reach earth's surface are the result of the atmosphere being bombarded by the energetic ionized nuclei that fill the interstellar medium of our galaxy. These nuclei have kinetic energies ranging from the GeV scale to billions of GeV~\cite{Abbasi2023}. Because of their many catastrophic nuclear and electromagnetic interactions in the atmosphere, the ground-level cosmic rays consist of a wide variety of particle types ($e^\pm$, $\mu^\pm$, $\gamma$, protons, and neutrons) carrying an enormous range of energies. The diversity of particle species and energies, and the possibility of generating secondary particles makes shielding instruments from cosmic rays a serious challenge.

\section{Simulation methodology} \label{sec:methodology}

\begin{figure*}[ht!]
    \centering
    \includegraphics[width=0.325\linewidth]{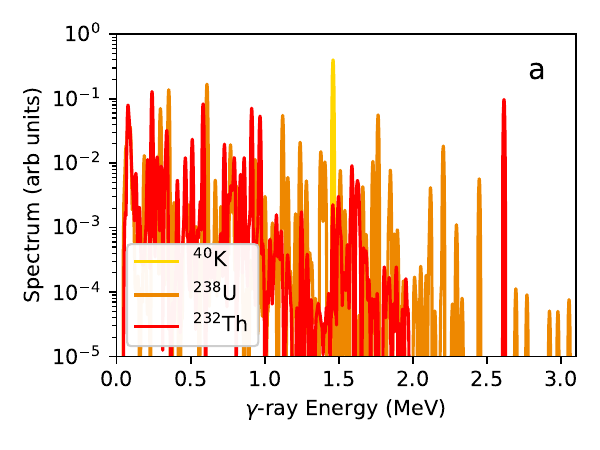}
    \includegraphics[width=0.325\linewidth]{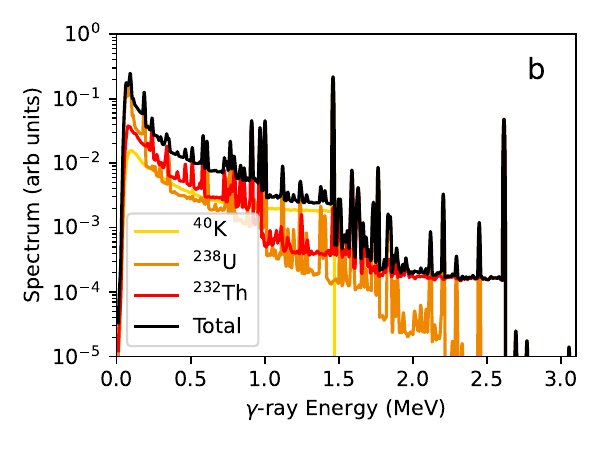}
    \includegraphics[width=0.325\linewidth]{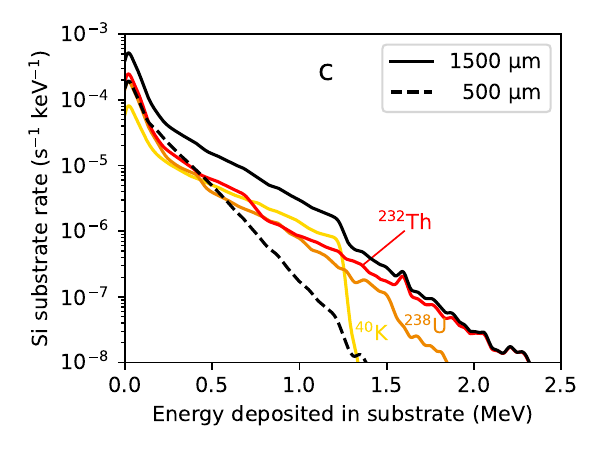}
    \caption{Spectrum of terrestrial gamma rays and x rays (a) as emitted by perfectly isolated radionuclides. (b) The same gamma rays, emerging from a thick slab of concrete, with decays homogeneously distributed throughout the slab. Both panels a and b have the y-axis in arbitrary units and assume the same relative activities as the nominal case. (c) The spectrum of energy deposited by the nominal activity levels in a silicon substrate of the nominal thickness \qty{500}{\micro\m} (dashed) or  \qty{1500}{\micro\m} (solid). The as-deposited spectrum (c) falls more rapidly with energy and lacks distinct lines, because thin silicon substrates in the gamma-ray band have nearly zero probability for photoionization, the mechanism that leads to capture of the full photon energy.
    }
    \label{fig:gamma-spectrum}
\end{figure*}

We have modeled cosmic rays and terrestrial gamma rays to characterize the energy deposited in wafers of silicon or other possible substrate materials. We have used the particle-transport Monte Carlo simulation Geant4~\cite{Geant2003,Geant2006,Geant2016} for the study of energetic particles in shielding and substrates. The simulations are configured by the Tool for Particle Simulation (TOPAS)~\cite{Perl2012,Faddegon2020}, a framework providing simple, text-driven control over the full complexity of Geant4.

We have run particle simulations in two separate steps, whether for terrestrial or cosmic sources. In the first, shielding is treated as simple, very broad two-dimensional slabs of material. For cosmic rays, this shielding includes both a concrete ceiling and a \qty{1}{cm}-thick aluminum layer to represent the effect of cryostat shells. For up-going terrestrial gamma rays, shielding includes the thick concrete foundation from which gammas are emitted, plus the same aluminum layer.

After the first step, outgoing particles are diverging in all directions from the initial impact point. The fraction that would then strike any centimeter-scale substrate would be counted in the parts per million, representing a hugely inefficient computation. Although one could simulate an enlarged substrate to overcome the inefficiency, this change would hide any edge effects---particles entering or exiting the lateral area of the simulated wafer. Instead, we have ``re-aimed'' particles after the first step: particle trajectories are shifted by an amount that uniformly strikes the cross-section of a small sphere centered on the substrate model, while their direction is preserved. This shift requires re-scaling the effective time being modeled by the ratio of the unshifted to the shifted areas, which is typically in excess of $10^7$.

We wish to study many variables: the substrate material, thickness, area, and shape, as well as the amount of shielding, and (for cosmic rays) the observing elevation. This represents a six-dimensional space. For efficient exploration, we have defined a nominal substrate, changing only one parameter to generate each alternative model. The \emph{nominal absorber} is:
\begin{itemize}
    \item A silicon wafer \qty{500}{\micro\m} thick,
    \item \qty{10}{mm}$\times$\qty{10}{mm} square,
    \item at sea level,
    \item shielded from all particles by \qty{1}{cm} of aluminum, 
    \item and from cosmic rays by \qty{20}{cm} of concrete roofing.
\end{itemize}
Any superconducting sensor or qubit will be constructed from a thin film. We ignore this film and consider only energy deposited in the substrate, which is typically far thicker and dominates the absorption of energy from background radiation.

Each simulation yields a spectrum of the energy deposited in the substrate. Although the entire spectrum is of some interest, we find it useful to summarize each by three key rates:
\begin{enumerate}
    \item The rate $R$ of events that deposit any energy, $E>0$.
    \item The rate $M$ of events where $E>$\qty{1}{MeV} is deposited.
    \item The total power $P$ deposited by all events.
\end{enumerate}
The threshold of \qty{1}{MeV} for the $M$ rate is chosen because it represents a transition from the low-$E$ part of the spectrum, which is dominated by terrestrial gamma events and by cosmic-ray electrons and muons, to the higher-$E$ events, which are dominated by cosmic-ray protons and neutrons striking the substrate. Furthermore, \qty{1}{MeV} represents the energy above which events in the nominal substrate occur once per hour. This specific rate is highlighted as a concern in~\cite{Acharya2024}.

\section{Terrestrial gamma-ray models and results}

Gamma rays are assumed to be generated uniformly throughout a slab of concrete \qty{50}{cm} thick. The attenuation length in concrete is \qty{11}{cm} or less for gamma rays of energies less than the \qty{3}{MeV} maximum from the \Uranium\ and \Thorium\ chains~\cite{xcom}, so a \qty{50}{cm} slab is a realistic proxy for any slab of at least that thickness. Figure~\ref{fig:gamma-spectrum} shows the gamma-ray spectrum as emitted (panel a) and above the concrete slab (panel b), as well as the spectrum as absorbed in the nominal silicon substrate (panel c). The numerous gamma-ray lines in the emitted spectrum (b) are absent from the deposited spectrum (c), because semiconductor substrates thinner than \qty{1}{mm} are optically thin to photoionization~\cite{xcom}. The rare interactions that occur are primarily Compton scattering, where only a fraction of the gamma-ray energy is deposited in the substrate.

\begin{table}[]
    \centering
    \begin{tabular}{lll}
    Parameter & Nominal & Alternatives \\ \hline
    Material & Si & SiC, SiO$_2$, Al$_2$O$_3$, GaN, GaAs \\
    Thickness (\unit{\micro\m}) & 500 & 1, 3, 10, 30, 100, 300 \\
    Size (mm$\times$mm) & $10\times10$ & $10\times\{5, 2, 1\}$, $7\times7$, $5\times5$ \\
    Ceiling (cm concrete) & 20 & 3, 40 \\
    Elevation (m a.s.l.) & 0 & 500, 1000, 1500, 2000, 3000
    \end{tabular}
    \caption{Nominal values of each parameter and alternatives studied. (Ceiling and elevation do not affect gamma-ray models.)}
    \label{tab:variable_params}
\end{table}

Table~\ref{tab:variable_params} summarizes the range of studies we performed, giving the nominal value of each variable and the alternatives used. In every model, no more than one parameter was changed from the nominal value. This approach to exploring the high-dimensional space means that we gain no insight into any correlated effects such as of higher elevation and thicker ceilings, or between higher-density substrates and long narrow shapes. We decided this was a minor sacrifice for the benefit of a smaller and simpler problem.
\begin{figure}
    \centering
    \includegraphics[width=\linewidth]{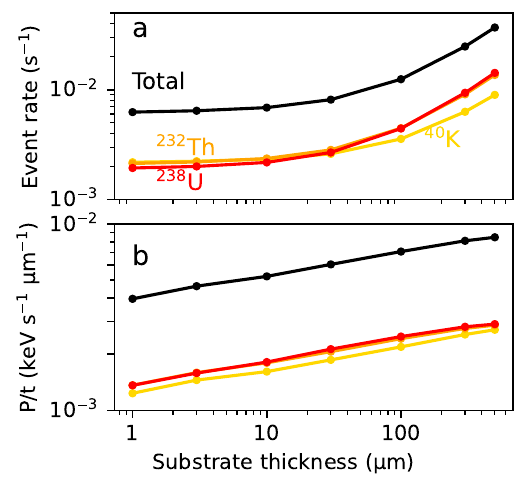}
    \caption{Dependence of terrestrial gamma-ray effects on substrate thickness $t$. (a) The total event rate in the nominal substrate. (b) The power deposited, per unit substrate thickness. The event rate scales linearly with $t$ for thicker substrates, while the power scales approximately as the 1.12 power of $t$.}
    \label{fig:gamma-thickness}
\end{figure}

The gamma-ray event rate $R_\gamma$ is proportional to wafer thickness for thicker wafers ($t\gtrsim \qty{100}{\micro\m}$) but approaches a non-zero constant for thinner substrates (Figure~\ref{fig:gamma-thickness}a). We attribute the proportional term to gamma rays, for which the substrate is nearly transparent, and the constant term to Compton-scattered electrons ejected from the concrete floor and the aluminum vacuum shell. Such electrons interact in even the thinnest silicon substrates. The power $P_\gamma$ grows approximately as  $t^{1.12}$ (Figure~\ref{fig:gamma-thickness}b), because thicker substrates not only cause more Compton-scattering events but also absorb more energy from the scattered electrons. The MeV-scale rate is difficult to estimate from the few simulated events above that energy, but $M_\gamma\propto t^5$ approximately captures the scaling.

\section{Cosmic-ray background models and results}

Modeling the cosmic-ray background radiation requires a random source of cosmic rays that reflects the distribution of energies and zenith angle for cosmic rays of each particle species. Many cosmic-ray models characterize only the muons, or apply only to a specific elevation. We use the model PARMA~\cite{Sato2013,Sato2016}, because it separately characterizes $p, n, \mu^+, \mu^-, e^+, e^-, \mathrm{and}\ \gamma$, and it captures changes with elevation.  It parameterizes the spectra with formulas fit to the results of a large library of simulated cosmic-ray air showers.

We sampled from the distributions and generated TOPAS ``phase space'' input files. As with the gamma-ray models, we ran the simulation in two Geant4 steps: first through shielding to represent a ceiling and cryostat (\qty{20}{cm} of concrete and \qty{1}{cm} of aluminum), then after re-aiming the resulting particles, we ran a second step to assess energy deposited in the substrate.

\begin{figure}
    \centering
    \includegraphics[width=\linewidth]{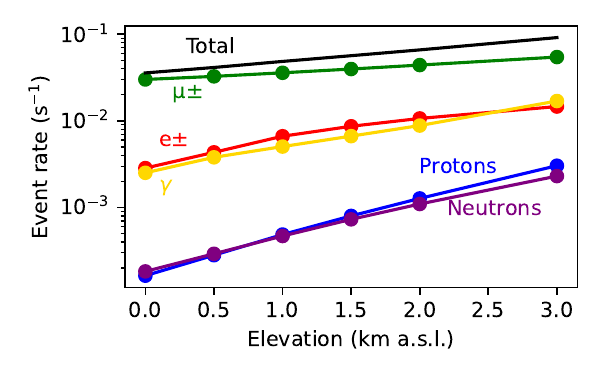}
    \caption{Elevation dependence of cosmic-ray event rate in the nominal substrate, in total and separated by primary particle species (that is, particle type before any ceiling or shielding). The scale length of electromagnetic particles ($e^\pm, \gamma$) is longer than that of nuclear particles but less than that of the $\mu^\pm$. The power deposited (not shown) scales similarly.
    }
    \label{fig:CR-altitude}
\end{figure}

\begin{figure}
    \centering
    \includegraphics[width=\linewidth]{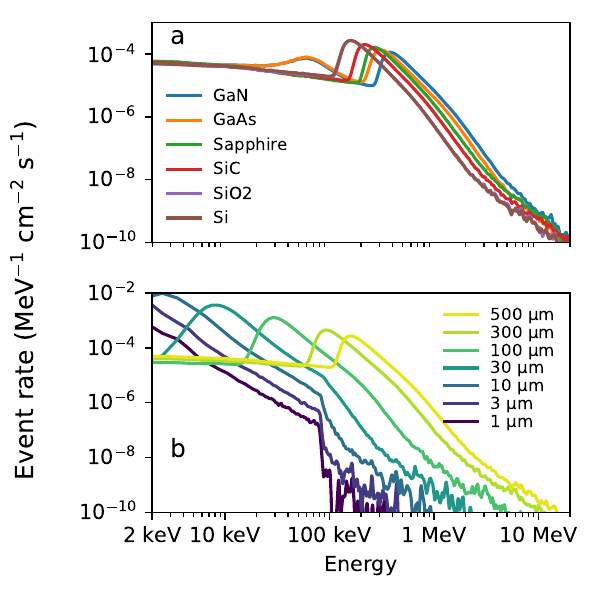}
    \caption{Spectrum of energy deposited by cosmic rays (all particle species) in $10\times$\qty{10}{mm^2} substrates: (a) for various substrate materials of the nominal thickness (\qty{500}{\micro\m}), and (b) for silicon substrates of various thickness.
    }
    \label{fig:CR-spectra}
\end{figure}

The cosmic-ray event rate grows exponentially with altitude, as expected~\cite{Sato2016}. The characteristic scale height in the atmosphere is not equal for all particle species, however (Figure~\ref{fig:CR-altitude}). The scale height is approximately \qty{5}{km} for muons, \qty{1}{km} for nuclear particles, and in between these values both for electromagnetic particles ($e^\pm, \gamma$) and for the overall event rate.

Figure~\ref{fig:CR-spectra} shows the spectrum of deposited energy as it depends on substrate material (panel a) and thickness (b). In all cases, a wide but clear peak can be seen, centered around \qty{170}{keV} for the nominal case. This peak is the result of the large population of minimum-ionizing charged particles passing through the full substrate thickness, albeit with a long tail due largely to lower-energy protons. The low energy losses reflect both gamma-ray photoabsorption and charged particles entering or exiting the lateral surface of the rectangular substrate. The former effect is visibly enhanced around \qty{100}{keV} for the substrates containing gallium.

\section{Summary of results}  \label{sec:summary}

\begin{table*}[]
    \centering
    \caption{Parameters of cosmic-ray and gamma-ray models, estimated from Geant4 simulations with PARMA spectra for the cosmic rays.}
    \label{tab:terrestrial-params}
    \begin{tabular}{crS[table-format=2.2]S[table-format=2.2]S[table-format=2.2]S[table-format=2.2]S[table-format=2.2]rr}
    Source & 
    \makecell{$n_s$ \\ (\unit{Bq.kg^{-1}})} & 
    \mc{\makecell{$c_s$ \\ ($10^{-3}$\,\unit{s^{-1}})}} & 
    \mc{\makecell{$g_s$ \\ ($10^{-3}$\,\unit{s^{-1}})}} & 
    \mc{\makecell{$p_s$ \\ (\unit{keV.s^{-1}})}} & 
    $\beta_s$ &
    \mc{\makecell{$m_s$ \\ ($10^{-6}$\,\unit{s^{-1}})}} & 
    $\alpha_s$ &
    \makecell{$\rho_{\mathrm{Ga},s}$ \\ (\unit{g.cm^{-3}}) } \\ \hline
    \Potassium & 400 & 2.2  &  6.8 & 1.4  & 1.12 & 15 & 5.0 &  2 \\
    \Thorium-a &  30 & 0.6  &  4.9 & 0.5  & $''$ & 2 & $''$ & 4 \\
    \Thorium-b &  30 & 1.5  &  6.6 & 0.9  & $''$ & 20 & $''$ & 4 \\
    \Uranium-a &  40 & 0.02 &  0.6 & 0.03 & $''$ & 0 & $''$ & 15 \\ 
    \Uranium-b &  40 & 1.9  & 11.7 & 1.4  & $''$ & 13 & $''$ & 4 \\
    Cosmic rays & --- & 40  & 1.4  & 8.0  & 1.0 & 180 & 1.8 & --- \\
    \end{tabular}
\end{table*}


In this section, we present as concise a summary as possible of the three rates across all simulations, while capturing their most important dependencies on the modeled conditions. A TKID-based spectroscopic measurement~\cite{fowler2024} is consistent with the models' results under nominal conditions (except that it was performed at \qty{1640}{m} above sea level).

The rates $R$, $M$, and $P$ are defined above (Section~\ref{sec:methodology}). All are proportional to the substrate area $A$. Minor adjustments for the size and shape of the substrates are required, generally less than \qty{5}{\%}. Each rate is a sum over six distinct sources of background emission: the five gamma-producing decay chains and cosmic rays (Eq.~\ref{eq:rates}). Each term in the sum is proportional to the relative activity $\tilde a_s$ of source $s$, compared to the nominal case. For the gamma chains, we define $\tilde a_s\equiv a_s/n_s$ where $a_s$ and $n_s$ are the actual and nominal specific activity of chain $s$ in the laboratory foundation. The cosmic-ray intensity depends on local conditions only as a growing function of elevation $H$, so we define the equivalent term for cosmic rays as $\tilde a_\mathrm{CR}\equiv\exp(H/\lambda)$ for an appropriate scale height $\lambda$.

The summary rates depend differently on the substrate's relative thickness $\tau\equiv t/\qty{500}{\micro\m}$. Each term in the event rate $R$ depends linearly on $\tau$, with both constants depending on the source. The power $P$ and the MeV-plus rate $M$ grow as different powers of $\tau$ for cosmic -ray and gamma-ray sources. We can express all three formulas in parallel:
\begin{equation} \label{eq:rates}
    \begin{bmatrix}
    R \\
    P \\
    M
    \end{bmatrix} = 
    \left(\frac{A}{\qty{100}{mm^2}}\right)
    \sum_{s}\, \tilde a_s
    \begin{bmatrix}
    c_s + g_s\tau\\
    \hspace{22pt} p_s \tau^{\beta_s} \\
    \hspace{22pt} m_s \tau^{\alpha_s} \\
    \end{bmatrix}
    \kappa_\mathrm{c} \kappa_\mathrm{sh} \kappa_\rho.
\end{equation}

Table~\ref{tab:terrestrial-params} gives the appropriate values for:
\begin{itemize}
    \item $n_s$, the nominal specific activity of each gamma-producing decay chain in typical concrete,
    \item $c_s$, the charged-particle event rate of source $s$ in silicon wafers of any thickness,
    \item $g_s$, the nominal gamma-ray event rate of source $s$,
    \item $p_s$, the power deposited by source $s$ in nominal wafers,
    \item $\beta_s$, the thickness scaling power law for $P$,
    \item $m_s$, the rate of $E>\qty{1}{MeV}$ events from source $s$ in nominal wafers, and
    \item $\alpha_s$, the thickness scaling power law for $M$.
\end{itemize}

The final factor in Eq.~\ref{eq:rates} is the product of three corrections for effect of the ceiling on cosmic rays ($\kappa_\mathrm{c}$, but $\kappa_\mathrm{c}\equiv 1$ for gamma rays), and for the substrate's shape ($\kappa_\mathrm{sh}$) and density ($ \kappa_\rho$). The ceiling and shape corrections are close to one for most circumstances; we include them as a way to discuss the sign and size of the effects.

The ceiling reduces $R$ and $P$ for cosmic rays by only \qty{2}{\%} per \qty{10}{cm} of added concrete. The shape correction depends on the relative area of the sides to the main surface, with the high-aspect-ratio $10\times$\qty{1}{mm^2} wafer showing a \qty{20}{\%} increase in $R$ and a  \qty{3}{\%} reduction in $P$. Too few MeV-scale events were generated to characterize any shape correction to $M$.

A density correction $\kappa_\rho$ is needed for substrates other than silicon. In general, the values in Equation~\ref{eq:rates} increase with higher relative densities, $\tilde\rho\equiv\rho/\rho_\mathrm{Si}$. The high x-ray photoionization cross section of gallium requires special care. For cosmic rays,  $P_\mathrm{CR}\propto\tilde\rho$, and we replace $g_\mathrm{CR}$ in the $R$ equation with \qty{.007}{s^{-1}} for gallium substrates (a $5\times$ increase). For gamma rays, both $R$ and $P$ scale approximately with relative density, but as if the true density for gallium substrates were increased by an amount $\rho_\mathrm{Ga}$ that depends on source, as given in Table~\ref{tab:terrestrial-params}. For both cosmic and gamma rays, the high-$E$ event rate scales roughly as $M\propto\tilde\rho^{2.7}$.

\section{Conclusion}
We have simulated the effects of cosmic and terrestrial radiation backgrounds in a variety of objects whose size, shape, and composition are representative of the substrates of superconducting qubits and superconducting sensors. We have modeled realistic quantities of radiation shielding and generated the expected spectrum of energy deposited in such substrates by both sources of background. We have summarized the spectra by formulas that capture how the relevant effects of radiation depend on substrate material and geometry and (for cosmic rays) on elevation. We believe that the absolute rates given in this work are accurate to some $\pm$\qty{25}{\%} over a range of realistic conditions, and that the values can be used productively for approximate but quantitative estimates of background levels. They can also give intuitive understanding of what shielding steps might benefit any specific experiment.

\section*{Acknowledgments}
We thank all our co-authors on \cite{fowler2024} for helping to set the direction of this modeling effort, and Matthew Nathale, Mark Keller, and Pieter Mumm for helpful comments on a draft of this work. We gratefully  acknowledge support from the U.S. Department of Energy, Office of Science, Office of Nuclear Physics, under Award Numbers DE- SC0021415 and DE-SC0023682.


\bibliographystyle{IEEEtran}
\bibliography{IEEEabrv,TKID_background}


 





\end{document}